\documentclass[aps,prb,twocolumn,showpacs,amsmath,amssymb,reprint]{revtex4-2}
\usepackage{dcolumn}
\usepackage{graphicx}
\usepackage{bm}
\usepackage{booktabs}

\begin{document}
	\title{In-Situ Manipulation of Superconducting Properties via Ultrasonic Excitation} 
\author{Biswajit Dutta$^1$$^,$$^2$, S.De$^2$$^,$$^3$, A. Banerjee$^2$}
\affiliation{$^1$Ecole Normale Sup´erieure de Lyon, CNRS, Laboratoire de Physique, F-69342 Lyon, France\\
	$^2$UGC-DAE Consortium for Scientific Research, University Campus, Khandwa Road, Indore-452001, India.\\
	$^3$Solid State and Structural Chemistry Unit, Indian Institute of Science, C. V. Raman Road, Bangalore-560012, Karnataka, India.}
\begin{abstract}
	We demonstrate in-situ manipulation of the critical temperature ($T_S$) and upper critical field ($H_{C2}$) of conventional and unconventional superconductors via ultrasonic excitation. Using an AC susceptibility measurements, we observed a reduction in  $T_S$  with increasing amplitude of the applied ultrasonic waves. This reduction exhibits a power law dependence on the excitation voltage, suggesting a non-linear coupling between the ultrasonic waves and the superconducting order parameter.  Analogous behavior was observed in cuprate superconductors, hinting at a possible link between the modified superconducting properties and the modulation of the antiferromagnetic network by ultrasonic excitation. Measurements on a paramagnetic material (Gd$_2$O$_3$) with quenched orbital angular momentum (L\,=\,0) revealed no change in magnetization even at extreme ultrasonic excitation amplitudes. This highlights the role of spin-orbit coupling in the observed effects and rules out the possibility of local temperature increases affecting the measurements. To further confirm this, we conducted auxiliary experiments where a Cernox temperature sensor was subjected to ultrasonic excitation, and the resulting temperature difference relative to a reference Cernox was recorded. Simultaneously, the power difference was measured to assess the impact of ultrasonic heating. This analysis revealed that localized heating effects become dominant above an ultrasonic amplitude of 10 V$_{pp}$. 
 \end{abstract}
\maketitle

 The magnetization (M), electric polarization (P) and strain tensor ($\epsilon$) are three fundamental important order parameters in condensed matter physics. The external parameters like magnetic field (H), electric field (E) and the force tensor ($\sigma$) to control M, P and $\epsilon$, respectively. In the multiferroic material the cross correlation between M and P is observed i.e. M can be controlled by H and E both, along with that P can also be controlled by E and H. Therefore because of this cross correlation the functionality of the material increases for device application. Similarly both M and P can be controlled by $\sigma$ in high pressure experiment, because the electronic functionality originates from the crystal structure with translational symmetry and application of external pressure disturbs the translational symmetry by disturbing the lattice. The hydrostatic expansion is expected to decrease the electron hopping and there by increase the electron lattice coupling, which resulting in a lower value of Curie temperature ($T_C$), where as the hydrostatic compression shows the opposite behavior of expansion \cite{c6r1}. So these observations suggest that change of $T_C$ depends on the type of structural change, i.e. hydrostatic compression or anisotropic compression accompanied by tensile and compressive strain \cite{c6r1}. Y. Moritomo et al.,\cite{c6r2} has reported in La$_{(1-x)}$Sr$_x$MnO$_3$ (0.15$\leq$x$\leq$0.5)  pressure stabilizes the ferromagnetic metallic state by enhancing the transfer interaction. La$_{0.7}$Sr$_{0.3}$MnO$_3$ is observed to destabilize the ferromagnetic metallic state along with  decrement of $T_C$ ($\sim$10\,K) while increasing the tensile strain \cite{c6r3}. The superconducting transition temperature ($T_S$) is also observed to increase with increasing the hydrostatic pressure in lanthanum based hole doped cuprate superconductor where as in case of electron doped Lanthanum based superconductor $T_S$ does not vary with hydrostatic pressure \cite{c6r4}. In case of a thin film the stretching (squashing) of CuO$_6$ octahedra enhances (decreases) T$_S$ \cite{c6z1,c6z2,c6z3,c6z4}, but this type of modulation is static and irreversible in nature hence difficult for technological application, because the deformation has to be reversible and dynamical as well for any kind of technological application. The possible solution for this is the ultrasonic shock wave. In this case the applied stress is function of time and results in a diffusive acoustic strain. Due to spin-orbit interaction and dipolar field the spins experience elastic deformation when the lattice is excited by an elastic wave, this type of coupling is known as magnetoelastic coupling (MEC) and the opposite effect  is known as magnetostriction (i.e. the lattice is deformed due to the change of magnetization) \cite{c6a1,c6a2,c6a3}. Therefore if the frequency and wave length of spin wave in a crystal lattice become comparable to the the frequency and wave length of the elastic wave then probability of MEC increases or hybridization of spin-lattice excitation (i.e magnon-phonon hybridization) is observed, this coupling phenomena between elastic and magnetic dynamic was first theoretically investigated almost six decades ago by C.Kittel \cite{c6a4}. This field has been attracted much more attention in recent times because of improved material growth technique and device fabrication methods \cite{c6a5,c6a6,c6a7,c6a8}.

 \begin{figure*}[t]
 	\centering
 	\includegraphics[width=15cm,height=12cm]{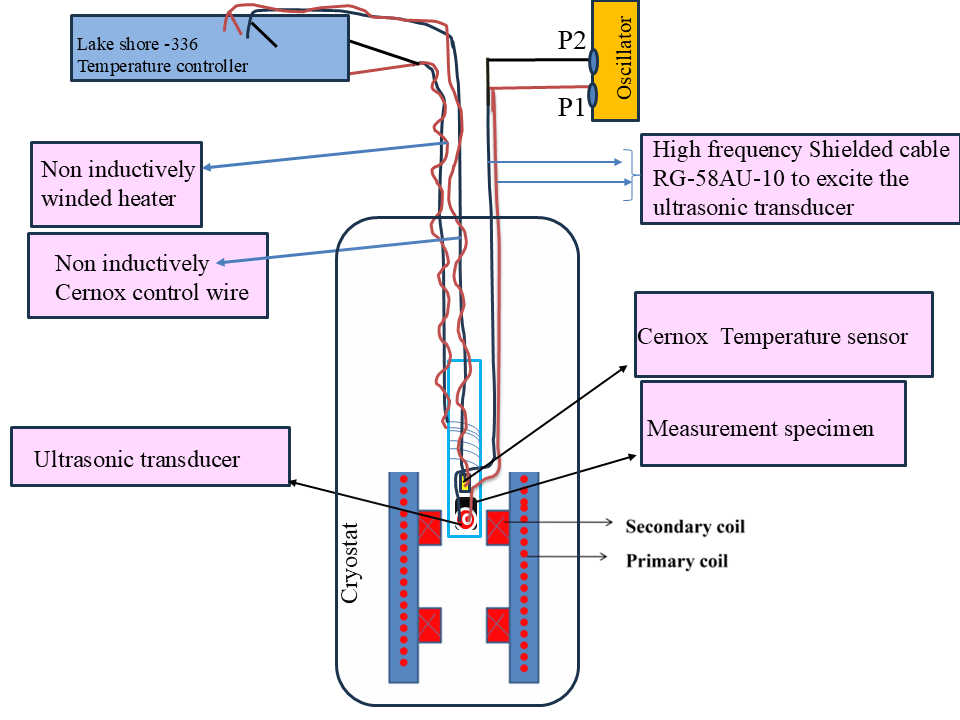}
 	\vspace*{0mm}
 	\caption{\textit{\small (Colour Online)The experimental arrangement for ulrasonic excitation during magnetization measurement. A detailed description of the experimental setup for ultrasonic excitation during magnetization measurements can be found in Ref\, \cite{r41}}}
 	\label{fig:fig2}
 \end{figure*}
  
 \begin{figure*}[t]
 	\centering
 	\includegraphics[width=15cm,height=12cm]{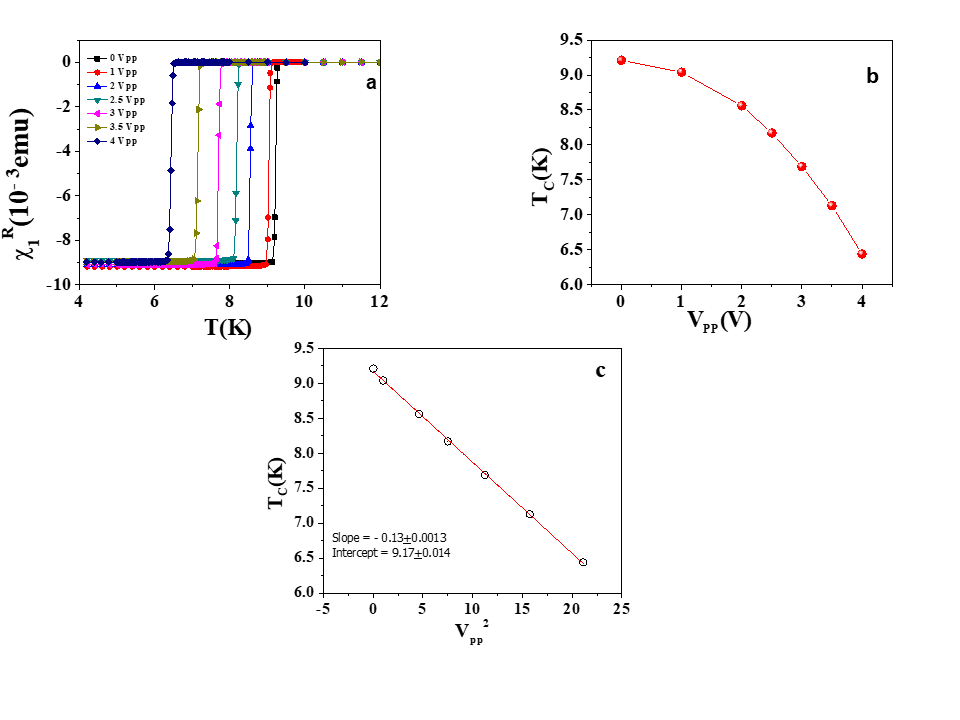}
 	\vspace*{-15mm}
 	\caption{\textit{\small(Colour Online) The measurements are performed on Nb, A "Y" cut, 10\,MHz LiNbO$_3$ crystal is used to excite the superconductor. The susceptibility measurements are performed in an ac field of amplitude 1\,Oe and frequency 231.1 Hz (a) The real part of first order susceptibility ($\chi_1^R$) is plotted against temperature and the corresponding V$_{PP}$ is 0, 1, 2, 2.5, 3, 3.5, 4\,V respectively.(b) T$_C$(K) is plotted against V$_{pp}$. (c) linearized plot of T$_C$(K) against V$_{PP}^2$.}}
 	\label{fig:fig2}
 \end{figure*}
 Recently in transverse spin Seebeck effect measurement acoustic wave has been used to pump spins in ferromagnet and MEC is concluded as the possible mechanism for the appearance of this effect \cite{c6a5,c6a6}. Ferromagnetic resonance of Cobalt film is also observed to get enhanced by pulsed surface acoustic wave \cite{c6a8}. The manipulation of Meissner effect of cuprate superconductor YBa$_2$Cu$_3$O$_7$ (YBCO) \cite{c6r5} and Bi$_2$Sr$_2$Ca$_2$Cu$_3$O$_{10}$(BSCCO(2223)) \cite{c6r6} is observed by dynamical acoustic strain, which indicates the importance of phonon contribution on the electron pairing mechanism of cuprate superconductor. The contribution of phonon to the superconductivity of cuprate superconductor is still under debate, the oxygen  isotope (O$^{18}$) effect measurement depicts the phonon contribution to the electron pairing mechanism is not simply follows the BCS theory \cite{c6i0,c6i1,c6i2,c6i3,c6i4,c6i5}. Previous Raman scattering studies by Liarokapis, Müller, Kaldis, and colleagues revealed a significant softening of the in-plane Ag oxygen mode in YBa$_2$Cu$_3$O$_x$  with increasing oxygen doping (x) \cite{r13}. This softening correlates with the rise in superconducting critical temperature (T$_C$) up to its maximum value, forming a dome-shaped relationship. Despite the critical role of oxygen doping in high-T$_C$ superconductivity \cite{r14}, this correlation remains an empirical observation lacking a clear theoretical explanation. Recent x-ray absorption spectra further demonstrate a striking link between anharmonic lattice displacements of Sr atoms (coupled to apical oxygens) and T$_C$ in YSr$_2$Cu$_{2.75}$Mo$_{0.25}$O$_{7.54}$ and Sr$_2$CuO$_{3.3}$\cite{r15}.These findings challenge unconventional explanations for high-T$_C$ superconductivity based on spin fluctuations or electron repulsion, highlighting the potential role of electron-phonon coupling.  Indeed, recent theoretical work \cite{r16,r17} has established a strong link between the softening of both optical and acoustic phonons and the underlying electron pairing mechanism. This suggests that lattice dynamics play a crucial role in high-T$_C$ superconductivity. 
  
  The millimeter-scale wavelength of the acoustic waves used in this study generates a long-range spatial modulation of the lattice, effectively modulating acoustic phonons. This modulation influences any property coupled to these phonons. Our primary objective is to demonstrate in situ manipulation of a superconductor's thermodynamic properties using this technique, while also shedding light on the electron pairing mechanism in  high-T$_C$ cuprates. To address potential concerns about localized heating, we replaced the sample with a Cernox sensor and monitored temperature changes with increasing excitation voltage.  No significant local heating was observed up to 10\,V$_{PP}$, as detailed in a subsequent section. 

\begin{figure*}[t]
	\centering
	\includegraphics[width=16cm,height=12cm]{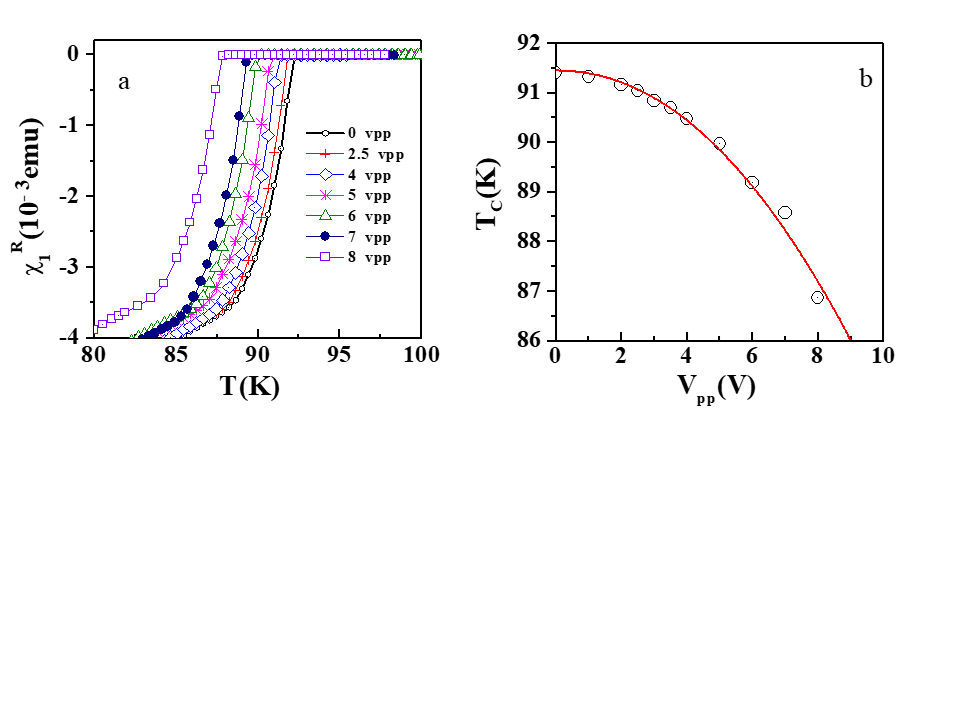}
	\vspace*{-55mm}
	\caption{\textit{\small(Colour Online) The measurements are performed on YBCO, A "Y" cut, 10\,MHz LiNbO$_3$ crystal is used to excite the superconductor. The susceptibility measurements are performed in an ac field of amplitude 1\,Oe and frequency 231.1 Hz (a) The real part of first harmonic ($\chi_1^R$) is plotted against temperature and the corresponding V$_{PP}$ is 0, 2.5, 4, 5, 6, 7, 8\,V respectively.(b) T$_C$(K) is plotted against V$_{pp}$.}}
	\label{fig:fig4}
\end{figure*}

The assembly of the experimental setup is shown in \texttt{FIG.\,1}. We have performed this measurement in a home built ac susceptibility set up\cite{r41}. It shows the concentric assembly of the primary and the secondary coil, a Lock-In amplifier (Sanford Research Systems model SR830) is used to excite the primary coil by the ac field, as well as for the measurement of the induced response from the secondary coils in differential mode. The temperature controlling and measurements are done by a temperature controller LakeShore model-336 and using Cernox sensor, respectively. The sample holder is made of single crystal sapphire rectangular bar of length 7\,cm and width 0.5\,cm and thickness of 0.2\,cm, this sapphire rectangular bar is grooved with a SS rod via a 10cm long hylum rod for better thermal isolation. A non inductively wounded heater wire (to minimize magnetic contribution from the heater current) and the Cernox temperature sensor is placed on the same sapphire bar. The 10MHz ultrasonic transducer made of LiNbO$_3$ is attached on the sample by very thin layer of stopcock grease and the transducer is excited from out side by using a high frequency function generator and high frequency cable (RG-58AU-10 of outer diameter 2.9\,mm) is soldered in the transducer for the electrical connection. Measurements are performed in the following protocols, initially the system is cooled to a desire temperature, then transducer is excited by an desirable 10\,MHz sinusoidal voltage and in this condition the temperature dependent ac susceptibility measurements are performed. Then after the final scan the transducer voltage is turned off and again the system has been cooled to the desirable temperature and theses process are repeated for all measurements.

Niobium (Nb) is a transition metal and shows conventional superconductivity below 9.25\,K(superconducting transition temperature (T$_C$)) \cite{c6r11} i.e. electron pairing is mediated by electron-phonon coupling (Follows BCS theory). Nb is mostly used in superconducting radio frequency cavity application \cite{c6r12}. The hydrostatic pressure dependent study on Nb depicted a decrement of superconducting transition temperature with the increment of hydrostatic pressure and change in temperature with the application of hydrostatic pressure follows the relation corresponding to -0.02\,K/GPa \cite{c6r13}. The superconducting property of Nb is also observed to be controlled by using epitaxial strain \cite{c6r14,c6r15,c6r16}. However these above approaches for the manipulation of phase transition and physical property of the superconductor are stationary or independent of time, hence these methods are not favorable for high speed manipulation in device application.

Here we have manipulated the phase transition by applying dynamical acoustic strain. Generally acoustic waves have been used to evaluate the elastic constant and sound velocity in any material, along with that the magnetic response of that system also modifies via MEC or spin-orbit coupling, electron-phonon coupling etc. However, we have used the piezoelectric transducer(LiNbO$_3$) as the source of time-dependent stress which acts as an external perturbation on the physical state of the superconductor. The temperature dependent ac susceptibility graph of Nb at different dynamical excitation amplitude (i.e. V$_{PP}$ of frequency 10\,MHz) is shown in \texttt{FIG.\,2a}, it shows the onset temperature of diamagnetism decreases to lower temperature when the amplitude of dynamical stress increases, here it is also observed that the transition remain sharper against temperature at maximum values of ultrasonic voltage, which indicates the excitation distributed uniformly throughout the bulk of Nb, other wise distribution of stress amplitude could have cause broadening in the superconducting transition temperature. At 4\,V$_{PP}$ the T$_C$ is observed around 6.25\,K, exhibits reduction almost about 3\,K. The variation of T$_C$ against V$_{PP}$ is shown in \texttt{FIG.\,2b}, it shows T$_C$ decreases by following the square of V$_{PP}$ i.e. the mathematical relation between T$_C$ and V$_{PP}$ is as follows,
\begin{equation}
T_C\,=\,9.22*[1-(\frac{V_{PP}}{7.31})^2]
\end{equation}

It describes the magnitude of ultrasonic strain on Nb is proportional to the power consumed by the transducer. It also depicts that at V$_{PP}$=\,7.31\,V, T$_C$ of Nb will become zero or electron pairing will not possible at this condition. The scaled voltage dependent curve of T$_C$ is shown in \texttt{FIG.\,2c}, which depicts a proper straight line graph with a negative slope, this observation also verifies the square law behavior of T$_C$ against V$_{PP}$. 

YBa$_2$Cu$_3$O$_7$ (YBCO) is a high T$_C$ cuprate superconductor and the electron pairing mechanism of this superconductor can not be explained by using conventional BCS theory or low energy phonon mediated electron pairing. \texttt{Fig.\,3a} shows the in phase ac susceptibility plot of YBCO at different amplitude of dynamical strain (i.e.V$_{PP}$). The sharp decrement of the diamagnetic fraction at V$_{PP}$=\,0\,V depicts good quality of the superconductor (i.e. the oxygen stoichiometry of every individual grain is almost same), along with that the similar sharpness of the decrement of diamagnetic fraction at maximum V$_{PP}$ indicates the equal distribution of the excitation over the whole volume of the pallet, it shows T$_C$ shifts to lower temperature with increase in the amplitude of V$_{PP}$. The dependence of T$_C$ on V$_{PP}$ is shown in \texttt{Fig.\,3b}. T$_C$ decreases proportionally to the square of V$_{PP}$ and the corresponding relation is given in \texttt{Eqn.\,2}

\begin{equation}
T_C\,=\,91.44*[1-(\frac{V_{PP}}{37.78})^2]
\end{equation}
The equation looks similar as \texttt{Eqn.\,1}, the only difference is the value of the voltage (i.e. V$_{PP}$=37.78\,V) at which  T$_C$ will reach to zero or YBCO will go to the normal state.
 \begin{figure*}[t]
	\centering
	\includegraphics[width=14cm,height=10cm]{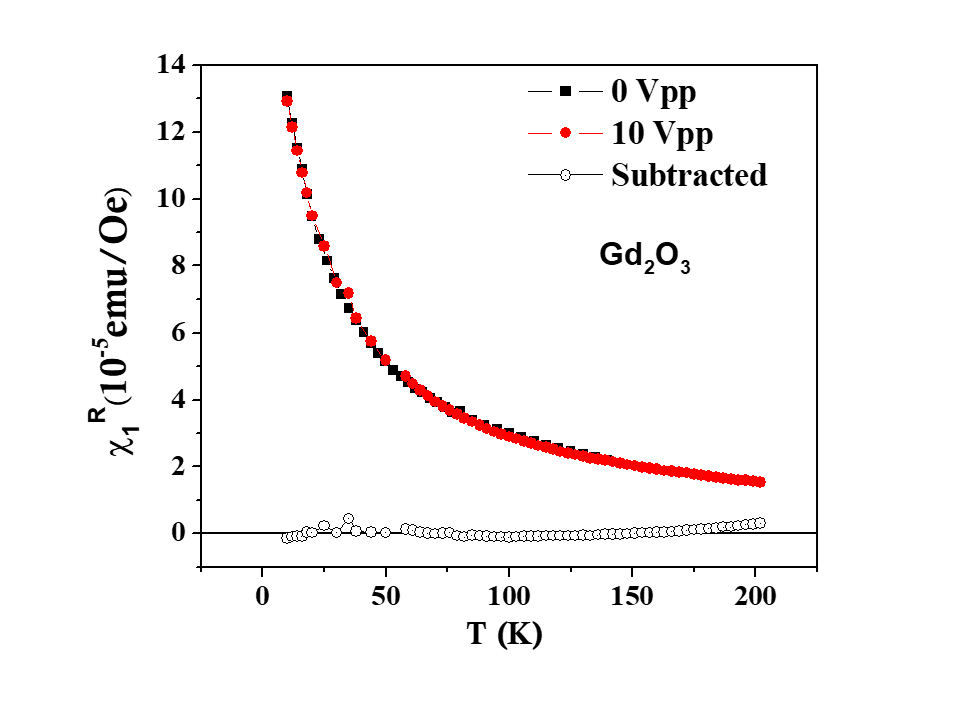}
	\vspace*{-6mm}
	\caption{\textit{\small(Colour Online) The measurements are performed on Gd$_2$O$_3$, A "Y" cut, 10\,MHz LiNbO$_3$ crystal is used to excite the paramagnet. The susceptibility measurements are performed in an ac field of amplitude 1\,Oe and frequency 231.1 Hz. The real part of first order ac susceptibility ($\chi_1^R$) is plotted against temperature and the perturbing ultrasonic voltage is V$_{PP}$ = 0\,V and 10\,V.}}
	\label{fig:fig5}
\end{figure*}
  
The real part of susceptibility (i.e. $\chi_{1}^R$) of Gd$_2$O$_3$ (GdO) at zero values of dynamic excitation amplitude (i.e. V$_{PP}$=0\,V) is shown in \texttt{Fig.\,4} along with that the susceptibility graph at  V$_{PP}$= 10\,V is also plotted. There is no change of the susceptibility value of GdO is observed at the maximum value of V$_{PP}$ also, obtained from the subtracted susceptibility plot (i.e. [$\chi_1^R$]$_{10\,V_{PP}}$\,-\,[$\chi_1^R$]$_{0\,V_{PP}}$). Which indicates that no change in magnetic susceptibility value even at maximum applied voltage (We have also performed the measurements at intermediate voltages, results are not shown for clarity), which indicates the Curie Weiss temperature ($\theta_C$) and effective magnetic moment ($\mu_{eff}$) remains same after exciting the system by ultrasonic excitation. The reason behind this is zero value of spin-Orbit (or spin-lattice) coupling. According to Russell–Saunders coupling scheme the ground state of $Gd^{+3}$ ions can be represented as $8_{S_{\frac{7}{2}}}$, depicting that $Gd^{+3}$ ion is isotropic in nature (or L=\,0), therefore  spin-orbit coupling is absent in GdO and spins only contribute to the magnetic property of GdO \cite{c6r17,c6r18}. Hence the result of GdO depicts the material having zero values of spin-orbit (or L-S) coupling is only going to show changes in the magnetic property while the system is excited by the ultrasonic excitation. As GdO has zero values of L-S coupling, due to this reason there is no change of the magnetic property is observed when GdO is excited by an ultrasonic excitation or acoustic wave.
 \begin{figure*}[t]
	\centering
	\includegraphics[width=14cm,height=6cm]{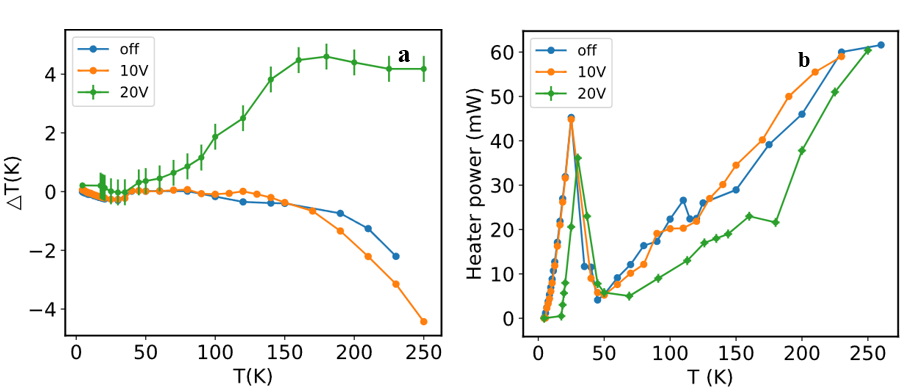}
	\vspace*{-1mm}
	\caption{\textit{\small(Colour Online) A "Y" cut, 10\,MHz LiNbO$_3$ crystal is used to excite the paramagnet. The temperature difference between two sensors (Cernox A and Cernox B) is measured at various temperatures. (b) The amount of heater power needed to reach and maintain those temperatures is also recorded and plotted against temperature.}}
	\label{fig:fig5}
\end{figure*}
To eliminate the possibility of localized heating effects induced by ultrasonic excitation, we conducted a series of control experiments. Our experimental setup involved mounting the sample on a single-crystal sapphire rod, chosen for its high thermal conductivity along the c-axis \cite{c6r19}. A non-inductively wound heater coil was attached to the top of the sapphire rod, followed by a Cernox temperature sensor (A) and the sample.

To generate ultrasonic waves in the sample, we used a transducer specifically designed to operate at the resonant frequency of the lithium niobate crystal (LiNbO$_3$ -10 MHz). This minimized energy loss within the crystal itself, ensuring efficient transmission of the ultrasonic signal \cite{r41a}. To accurately monitor any temperature changes caused by the ultrasonic waves, a second temperature sensor(Cernox B) was placed in contact with the transducer in the sample place as shown in \texttt{Fig.\,1}. This allowed us to compare any temperature variations at the sample location with the temperature measured by the primary sensor (Cernox A) with the applied voltage.

\texttt{Fig.\,5a} (blue graph) shows the temperature difference ($\Delta$T) between Cernox (B) and (A) as the sample space temperature was varied with the transducer off, establishing the calibration error between the two sensors. Subsequently, the transducer was activated at 10\,V$_{pp}$ amplitude, and the temperature-dependent $\Delta$T was measured (\texttt{Fig.\,5a}, orange graph). Each measurement point represents the average of 5 trials, with error bars indicating the range of observed values. Thermal equilibrium, as indicated by a constant $\Delta$T, was achieved within 4 minutes with ultrasonic excitation and the same time without, generally if there is any local heating that will also be stabilized with in this time\cite{r41a}. A five-minute stabilization period followed each ultrasonic excitation before susceptibility measurements, for safety. The temperature difference between A and B cernox  for 20\,V$_{pp}$ is very abrupt, which indicates a huge local heating effect, The size of the error bars reflects the difficulty in achieving a perfectly stable temperature and the range of fluctuation.

The power required to stabilize the sapphire bar at different temperatures is shown in \texttt{Fig.\,5b}. We recorded the power level only after the temperature reached a steady state for a specific configuration of the PID controller. Notably, the power needed remained consistent, within the error bar, regardless of transducer state (on at 10\,V$_{pp}$ or off). But the power required to stabilize a particular temperature for 20\,V$_{pp}$ voltage is consistently lower in amplitude for all temperature and it is also very difficult to stabilize the temperature when 20\,V$_{pp}$ voltage is applied.The fluctuations in power values (for 20\,V$_{pp}$), as indicated by the error bars, are a result of inconsistent heating generated by the high voltage within the transducer. The PID configuration remained constant throughout the experiment, demonstrating that the heating effect from the transducer voltage alone is responsible for the observed effects on temperature stability and power needs.

The graph displays in \texttt{Fig.\,5b} depicts the relationship between temperature and the power needed to keep the sapphire bar at a stable temperature. Our experiment takes place in a vacuum environment around 10$^{-5}$mbar, but we use liquid helium to cool the sapphire bar below 50 Kelvin (50\,K).  Because some heat also escapes into the surrounding helium bath, we need to increase the power to keep the bar at a specific temperature below 50\,K. This explains the sudden rise in power needed at lower temperatures shown in the graph \texttt{Fig.\,5b}. Interestingly, when the transducer is off or operating at a lower voltage (10\,V), the power required is nearly identical. Maintaining a consistent temperature in the sapphire bar becomes more difficult when the transducer operates at a higher voltage (20\,V), despite requiring less power overall.The likely reason for the unstable temperature and lower power requirement at higher voltages (20\,V) is that the increased voltage causes localized heating within the crystal, and this heat then affects the entire experimental setup. To avoid this, we always use an operating voltage below 10\,V.  This may be due to the negligible tangent delta (tan$\delta$) till 10\,V$_{pp}$ voltage \cite{r41a}.
 
To verify our observations, we calculated how long it should take to heat the sapphire bar by one degree Kelvin (from 85\,K to 86\,K) when applying 10 mW of power at 85K. This calculation assumed no heat loss to the surroundings, meaning all the power goes directly into heating the sapphire.
The volume of the sapphire bar is$=$ 5 $\times$ 0.25 $\times$ 0.5 $\times$ 10$^{-6}$ \, $\text{m}^3$,
Density of sapphire =3980 kg/m$^3$,
So, m=0.0024875\,kg,
C= Specific heat of sapphire around 80\,K is =94.9$\times$ 10$^{-3}$J/kg$\cdot$K\cite{c6r19}.
So to increase the temperature from 85\,K to 86\,K the required amount of energy is = $\bigtriangleup$Q$=$m$\times$C$\times$$\bigtriangleup$T$=$0.0024875$\times$94.9$\times$1$=$0.23\,J
The amount of power we have provided is 16\,mW =0.016\,J/sec
So amount of time required to reach from 85\,K to 86\,K is =0.23/0.016=14.37\,sec

Which is very close to the time we have observed for a particular PID setting (which we use from 50\,K to 120\,K temperature range). We have used the thermalization time around 5 min after applying the ultrasonic voltage a typical time to stabilize a temperature in our system also the transducer temperature due to voltage excitation \cite{r41a}.Our findings indicate that the observed decrease in the critical temperature of both niobium (Nb) and yttrium barium copper oxide (YBCO) superconductors is not caused by localized heating from the ultrasonic transducer due to intrinsic loss because of the applied sinusoidal voltage. Instead, a different phenomenon, inherent to the nature of both Nb and YBCO, appears to be responsible. This phenomenon amplifies the internal vibrations of the atoms (phonons), effectively increasing their temperature. This suggests an intrinsic mechanism is at play, going beyond simple thermal effects from the transducer.

Nb is a conventional superconductor, hence the cooper pairing is mediated by low energy phonon, due to this reason while the system is excited by the ultrasonic wave then electron scattering enhances due to the modulation of phonon frequency and hence the probability of electron phonon coupling reduces. Due to this reason the T$_{S(onset)}$ shifts towards lower temperature with increasing the amplitude of ultrasonic voltage. 

In case of YBCO, T$_C$ is also observed to decrease with increasing the amplitude of ultrasonic voltage in the similar manner as observed in case of Nb, indicates the involvement of low energy phonon in the electron pairing mechanism of cuprate superconductor, which is quite debated till today \cite{c6i0,c6i1,c6i2,c6i3,c6i4,c6i5}. In case of Nb, 4\,V$_{PP}$ sinusoidal voltage is required to reduce the T$_C$ about 3\,K, where as in case YBCO almost 7\,V$_{PP}$ sinusoidal voltage is required to reduce the temperature about 3\,K, which indicates the coefficient (also the nature) of electron phonon coupling in both the superconductors are different. These difference can be compared with the isotope effect in superconductor. According to BCS theory the shift in superconducting transition temperature (T$_C$) with isotope mass (M) decides the  electron phonon coupling coefficient of any superconductor \cite{c6i0,c6i1}, 
\begin{equation}
	\frac{\delta\,T_{C}}{T_C}=\alpha\,\frac{\delta\,M}{M}
\end{equation}
Where $\alpha$ is partial isotope shift factor \cite{c6i0}. In the weak coupling limit T$_C$ is given by 
\begin{equation}
	T_{C}\,=\,1.13\theta_{D}\exp{(-\frac{1}{\lambda})}
\end{equation}
$\lambda$ is electron phonon coupling constant and weakly depends on M, where as $\theta_{D}$ is Debye temperature and proportional to M$^{-0.5}$. The different isotope mass simply changes the Debye frequency, which affects the value of T$_C$ \cite{c6i0}. In case of conventional superconductor the value of $\alpha$ is observed around 0.5, whereas in case of cuprate superconductor this number is observed to vary from 0.03 to 0.16 \cite{c6i1}. There are several controversies in the existing literature regarding the involvement of phonon in the cooper pairing mechanism of high temperature cuprate superconductor \cite{c6i1,c6i2,c6i3,c6i4,c6i5,r15,r16,r17}. The angle resolved photo emission spectroscopy (ARPES) measurement depicted, the O$^{18}$ isotope doping affect the system in a particular energy window which lies in the range of 100-300 meV \cite{c6i3} and these energy window is observed to coincide with the energy range of J and 2J, where J is the super-exchange interaction of neighboring copper spins \cite{c6i3,c6i4} (i.e. the modulation of Debye frequency also modulates the AFM network in cuprate superconductors). The ultrasonic wave is an acoustic wave and its wavelength is of the millimeter order, hence resulting in the long-ranged spatial modulation of the lattice and this modulation can also effect the antiferromagnetic network in the CuO$_2$ plane by spin-lattice (or spin-orbit) coupling. So the reduction of T$_C$ due to ultrasonic excitation indicates the modulation of antiferromagnetic network (J) of the CuO$_2$ plane. Our previous experiments on ferromagnet and superconductor composite has also depicted the similar conclusion, in that work we have observed that the superconductivity of the cuprate superconductor is effected when the AFM interaction between the Cu$^{+2}$ spins are modulated due to the exchange field of the ferromagnet \cite{r42}. So, further studies are required on ultrasonic excitation induced change of the magnetic property of  superconductor having different hole concentration, because the AFM correlation length of cuprate superconductor is different in different hole doped region. Hence, this type of study can also provide a path way to find out the correlation between AFM fluctuation and superconductivity in the cuprate superconductor.

This study establishes ultrasonic excitation as a novel and effective method for manipulating both orbital and lattice excitations (phonons) in materials, enabling in-situ control of these crucial degrees of freedom.  The impact of this technique was confirmed through observations of  phonon-driven changes in the magnetic properties of Gd$_2$O$_3$, a paramagnetic salt chosen for its lack of orbital angular momentum.  Crucially, the absence of local heating effects from ultrasonic excitation was verified, ensuring the observed changes were intrinsic to phonon modulation.

Furthermore, we successfully demonstrated the manipulation of the energy gap in a conventional superconductor (Nb) using this method, opening exciting possibilities for applications in superconducting quantum devices.  Our findings also pave the way for new investigations into the role of phonons in high-Tc cuprate superconductors, suggesting a link between ultrasonic excitation, phonon modulation of the antiferromagnetic network, and the observed reduction in superconducting transition temperature.This technique offers a compelling alternative to conventional methods for manipulating critical thermodynamic parameters in superconductors, with advantages in controllability and time efficiency.  Potential applications include decoupling superconductivity from two-level systems (TLS) in qubit-based experiments, replacing persistent switch heaters in superconducting magnets, and other scenarios requiring precise and dynamic tuning of superconducting properties. This work highlights the broad utility of ultrasonic excitation as a tool for manipulating and tuning material properties, with significant implications for both fundamental research and technological applications.

This paper is dedicated to the memory of the late Dr. Alok Banerjee. The authors would also like to acknowledge Er. Kranti Kumar Sharma for insightful discussions.
 
\end{document}